\documentclass[a4paper,11pt,openany]{article}
\usepackage{graphicx}
\usepackage[applemac]{inputenc}
\usepackage{t1enc}
\usepackage{lscape}
\usepackage{amsmath}
\usepackage{amssymb}
\usepackage{cite}
\usepackage{epsfig} 
\usepackage{ulem}
\usepackage [dvipsnames] {xcolor}

\usepackage{xcolor}
\newcommand{\REVISED}[1]{{#1}}     

\usepackage[colorlinks=true,linkcolor=blue]{hyperref}

\begin{document}

\begin{center}
\Large{\bf Temporal reversibility of a fluid mixture under concentration gradient}
\end{center}

\begin{center}
O. Politano$^{a}$, Alejandro L. Garcia$^{b}$, F. Baras$^{a}$,  and M. Malek Mansour$^{c\dag}$
\end{center}

\begin{center}
\small{ (a) Laboratoire Interdisciplinaire Carnot de Bourgogne, $\quad$\\
 UMR 6303 CNRS-Universit\'e Bourgogne Franche-Comt\'e,\\
 F-21078 Dijon Cedex, France

(b) Dept. Physics and Astronomy,\\
 San Jose State University, \\
San Jose, California, 95192 USA

 (c\dag) Universit\'e Libre de Bruxelles CP 231, Campus Plaine,\\
B-1050 Brussels, Belgium (deceased)}
\end{center}

\date{\today}

\begin{abstract}

A binary fluid mixture in contact with lateral particle reservoirs is considered. By imposing different particle concentrations in these reservoirs, the system can be maintained under controlled non-equilibrium conditions. Previous stochastic approaches have revealed an unexpected property of the system's state trajectory, namely that it remains time-reversible even when the system is driven out of equilibrium. In the absence of relevant experimental evidence, we employ microscopic molecular dynamics simulations to assess the validity of this surprising result. Remarkably, the simulation results unambiguously confirm the prediction of the stochastic analysis.
 
\end{abstract}

\section{Introduction}\label{IntroSection}   


The time irreversibility of a state trajectory (or sample path) in physico-chemical systems \REVISED{is an indicator} of their non-equilibrium nature \cite{Mazur:1984,Kondepudi:2008}. However, \REVISED{to our knowledge}, the converse does not necessarily hold: a state trajectory may be time-reversible in a stationary regime even when the system is maintained out of equilibrium. A notable example is provided by single-variable Schl\"{o}gl-type reactive systems, in which reactions proceed either as  $X \rightarrow X+1$ (forward) or $X \rightarrow X-1$ (backward), where $X$ is the number of reactive particles\cite{Schl1:1971,Schl2:1972}.

This seemingly paradoxical property of time reversibility in non-equilibrium systems can be rigorously demonstrated within the standard Markovian stochastic framework used to model reactive systems \cite{Graham:1971,VanKampen:1983} (for a simple illustration, see Section 6.3 of \cite{Gardiner:2009}). Extensive numerical studies, including exact hard-sphere molecular dynamics simulations, have unequivocally confirmed these theoretical predictions for homogeneous reactive systems \cite{Baras:2023,Politano:2024}.

This raises a natural question: is this result confined to specific categories of reactive systems, or could similar behavior also arise in other non-equilibrium systems? Here we explore this issue by investigating the case of non-reactive fluid mixtures subjected to concentration gradients.  The theoretical analysis of such systems can be carried out using the master equation approach, in which particle motion is modeled as a random walk. In this work, we adopt the same framework that has been widely used to study the statistical properties of reaction-diffusion systems \cite{N-P:1977}.

An unexpected consequence of this formulation is that it yields time-reversible sample paths in the stationary regime, even when the system is maintained far from equilibrium. This outcome can plausibly be attributed to the highly simplified formulation of diffusion as a random walk. As such, a more thorough investigation is therefore required, including rigorous microscopic simulations (e.g., molecular dynamics), which require substantial computational resources. As we show, results from such molecular dynamics simulations indicate that, indeed, the sample paths are time-reversible.   \REVISED{It is instructive to recall that numerous articles have been devoted to this subject and that all reach the same conclusion: "forward" sample trajectories always differ from "reverse" sample trajectories in non-reactive systems maintained out of equilibrium. Our observation therefore contradicts all existing theoretical work.  A good article summarizing this fact before $1992$ is that of  J.A.G. Roberts and G.R.W. Quispel, \cite{Rob-Q} (see Refs \cite{Bull,Wil}).  Many others have since appeared, among which we may mention a few that are relatively well cited \cite{Ev-Coh,Ga-Coh,jarki,crk,Lebo,Sei}.}.

\section{Stochastic approach}\label{StochasticSection} 

Consider an ideally dilute, nonreactive fluid mixture confined within a box of dimensions $L_{\rm x}, L_{\rm y}$ and $L_{\rm z}$, in contact with two particle reservoirs located at $X = 0$ and $X = L_{\rm x}$. Periodic boundary conditions are imposed in the $Y$ and $Z$ directions. For simplicity, we focus on the diffusion of a single species; extension of the analysis to multicomponent systems is straightforward.

The system is partitioned into $N$ cells, labeled from $1$ to $N$, containing $X_1, \dots, X_N$ particles, respectively. The boundary cells, labeled $0$ and $N+1$, represent the reservoirs and contain fixed particle numbers: $X_A$ in the cell $0$ and $X_B$ in the cell $N+1$. Let ${\bf X} = (X_1, \dots, X_N)$ denote the state of the system. The dynamics is modeled as a simple random walk process.  The corresponding master equation reads:
\begin{align}
\label{eq1}
\frac{d }{d t}  P({\bf X},  \, t) &= D  \bigg\{ X_A \Big[P(X_1 - 1,\cdots, t)  \, -  \,  P({\bf X},  \, t)  \Big] + (X_1 + 1) \, P(X_1 + 1,\cdots,   t)   -  X _1\,  P({\bf X},  \, t)  \nonumber \\
 &+ (X_1 + 1) \, P(X_1 + 1,  X_2 -1,\cdots,  t)   -  X _1\,  P({\bf X},  \, t)  \nonumber \\
 &\vdots  \\
 &+  (X_N + 1) \, P(\cdots, X_{N-1} - 1,  X_N + 1,   t)   -  X _N\,  P({\bf X},  \, t)  \nonumber \\
 &+ \ X_B \Big[P(\cdots, X_N - 1,  t)  \, -  \,  P({\bf X},  \, t)  \Big] + (X_N + 1) \, P(\cdots, X_N + 1,   t)  -  X _N\,  P({\bf X},  \, t)  \bigg\}. \nonumber
\end{align} 
where the ellipsis ("$\cdots$")  denotes the variables $X_i$ that are not involved in the current particle-exchange step. As shown in \cite{Chris:1977}, Eq.  (\ref{eq1})  can be solved exactly for arbitrary initial conditions. Simple, though lengthy and tedious, calculations demonstrate that the resulting sample paths are time-reversible in the stationary regime, even when $X_A \ne X_B$ \cite{Turner:1977}. 

Given the unexpected nature of this result, one could begin by presenting the essential elements of the proof. However, rather than trying to summarize the corresponding lengthy derivation, we instead focus on the existence of time-reversible trajectories in such a simple system. \REVISED{We note that the time-reversibility of non-equilibrium state trajectories can be readily verified by simulating Eq. (\ref{eq1}) using the well-known standard Gillespie algorithm. \cite{Gil1,Gil2,Gil3}}. In particular, one may wonder whether this unusual property is a general feature \REVISED{of systems where the diffusion coefficient depends on the state of the system}, or whether it is restricted to stochastic models, such as those described by the master equation (\ref{eq1}). In the absence of relevant experimental data, the most natural alternative is to resort to classical molecular dynamics simulations.

\section{Microscopic simulation}\label{MolDynSection}

\subsection{Molecular dynamics formulation}

We consider a fluid mixture composed of two species, $A$ and $B$. As in the case of the stochastic modeling approach, we assume that they are confined in a cubic box of dimensions $L^3$, in contact with two reservoirs located at $X = 0$ and $X =L$.   Periodic boundary conditions are imposed in the $Y$ and $Z$  directions.  For convenience, in our microscopic simulations, the particles are treated as hard spheres of equal mass $m$ and radius $r$. 

When a particle emerges from one of the reservoirs, its identity is assigned consistently with the composition of that reservoir. By prescribing different compositions for the left and right reservoirs,  the desired non-equilibrium conditions (i.e., a concentration gradient) can be imposed on the system.  There are a variety of ways to implement reservoir boundary conditions in molecular dynamics. A common numerical method is to use perfectly rigid walls, such that the interaction between particles and the walls reduces to simple elastic collisions. However, a particle's identity is reassigned to match that of the corresponding reservoir.

A key advantage of this technique is that the presence of reservoirs does not modify the system dynamics, which remain those of a closed system. Consequently, the system \REVISED{eventually} reaches a steady state corresponding to thermodynamic equilibrium, whose properties do not interfere with the non-equilibrium features arising from the labeling of particles as species $A$ or $B$.

\begin{figure}
    \centering
    \includegraphics[width=0.70\linewidth]{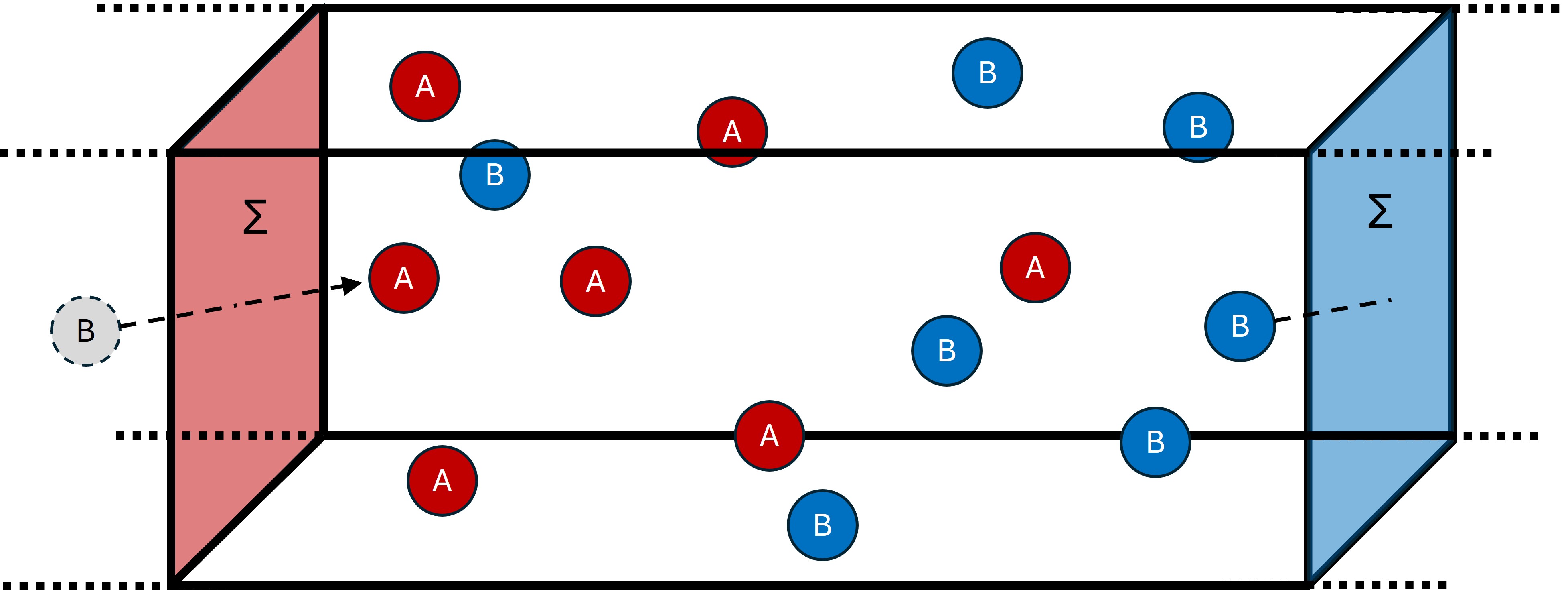}
    \caption{Illustration of a simple reservoir model for molecular dynamics. Here, as in scenario I, the left reservoir concentration is 100\% species $A$ so a particle of species $B$ crossing the periodic boundary $\Sigma$ from left to right has its identity set to species $A$. The solid heavy lines demark the system volume and the dotted heavy lines indicate the periodic boundary extension.
}
    \label{fig:MalekBox}
\end{figure}

In this work, we adopt an even simpler approach while preserving the advantages of the traditional method described above. We impose periodic boundary conditions and account for the presence of the reservoirs by introducing a fictitious surface, denoted $\Sigma$, at an arbitrary position within the system.   When a particle crosses $\Sigma$ from left to right, its identity is assigned to correspond to that of the left reservoir. Conversely, when it crosses the surface from right to left, its identity is modified in accordance with the properties of the right reservoir (see Figure~\ref{fig:MalekBox}).

Fundamentally, there is no difference between the two approaches. However, from a purely computational standpoint, implementing particle-wall interactions is more demanding than allowing particles to cross a fictitious surface. As a result, the execution of the numerical simulations becomes slightly faster. Given the enormous amount of data required to achieve our objective, this simplification is far from negligible.

Furthermore, the unusual properties highlighted in the previous section (Section~\ref{StochasticSection}) are obviously associated with events that occur over two distinct time intervals. They are therefore governed by two-body probability distribution functions. The generation of statistically reliable data for estimating two-body probability distribution functions constitutes a major challenge, as it requires a very large number of data points. One way to mitigate this problem is to perform molecular dynamics simulations of relatively dilute hard-sphere systems, in which careful management of the collision schedule can substantially enhance computational efficiency \REVISED{(see \cite{Allen:2017, Frenkel:2001, Rapaport:2004})}.

Accordingly, for a system containing a total number of $N$ particles, we fix the volume so as to obtain a number density of $3 \times 10^{-3}$ particles per cubic diameter $d^3$. This choice ensures that the system remains within the regime of validity of the Boltzmann equation \cite{Bird:1976}. We note that the same choice was adopted in our previous work on the microscopic simulation of reactive systems \cite{Politano:2024}.

However, this algorithmic improvement alone proves insufficient because of the large number of configurations that must be analyzed.  These configurations are determined by the spatial distribution of species $A$ and $B$ within the system. After several unsuccessful attempts, we ultimately chose to focus on global quantities, such as $X_A$ and $X_B$, which denote the total numbers of species $A$ and $B$, respectively, and $X_{Tot} = X_A + X_B$, the total particle number. Since $X_{Tot} = N$ remains constant, the evolution of the system is governed by a single variable, either $X_A$ or $X_B$; somewhat arbitrarily, we chose to focus on $X_B$.

\section {Molecular dynamics results}

For the microscopic simulations, we consider a mixture of species $A$ and $B$ with $N$ total particles.  We start with the simplest case (scenario I), where each reservoir contains particles of only one species: species $A$ on the left and species $B$ on the right. Thus, any particle that crosses the surface $\Sigma$ from left to right is converted into species $A$, while those that cross it from right to left are converted into species $B$.
 
We then examine a less restrictive configuration in which the left reservoir consists of 80\% species $A$ and 20\% species $B$, while the right reservoir has the opposite composition (scenario \rm{II}). The comparison between these two scenarios is particularly useful for ruling out misleading interpretations and identifying potential programming inaccuracies.
  
Our primary objective is to analyze the statistical properties of stationary sample trajectories of the total number of $B$ particles, $X_B$, for both scenarios I and II.  By measuring $X_B$ over a sufficiently long time interval, we can estimate the "direct" probability distribution $P\big( X_{B}^{(1)}\!\!, t_1 \, ; \, X_{B}^{(2)}\!\!, t_2\big)$ corresponding to the presence of $X_{B}^{(1)}$ particles of species $B$ at time $t_1$ and $X_{B}^{(2)}$ particles at time $t_2$, with $t_2 \, > \, t_1$. We then compare the result with the corresponding "reverse" probability distribution $P(X_{B}^{(2)}\!\!, t_1 \, ; \, X_{B}^{(1)}\!\!, t_2)$ of having $X_{B}^{(2)}$ particles of species $B$ at time $t_1$ and $X_{B}^{(1)}$ particles at time $t_2$.  Special care is taken to ensure that the system has reached the stationary regime prior to data collection. Therefore, the resulting (estimated) probability distributions depend only on the time interval $\tau = t_2 - t_1$.

We observe that the statistical errors associated with these probability distributions increase rapidly for time intervals $\tau$ greater than 20 mean collision times (MCT). Consequently, we restrict our simulations to values of  $\tau$ not exceeding 20~MCT. For practical reasons, we fix the value of  $X_B$  at the final time $t_2 = t_1 + \tau$ to a "reference state" ${X}_B^\mathrm{ref}$.  As expected, the statistical error is smaller when the chosen value of $X_B^\mathrm{ref}$ is close to the most probable state.  Since the latter coincides with the macroscopic steady state, $X_\mathrm{ss}$, we set ${X}_B^\mathrm{ref} = X_\mathrm{ss} = (X_A + X_B)/2  = N/2$.

\begin{figure}[ht!]
\begin{center}
\epsfclipon
\epsfig{file=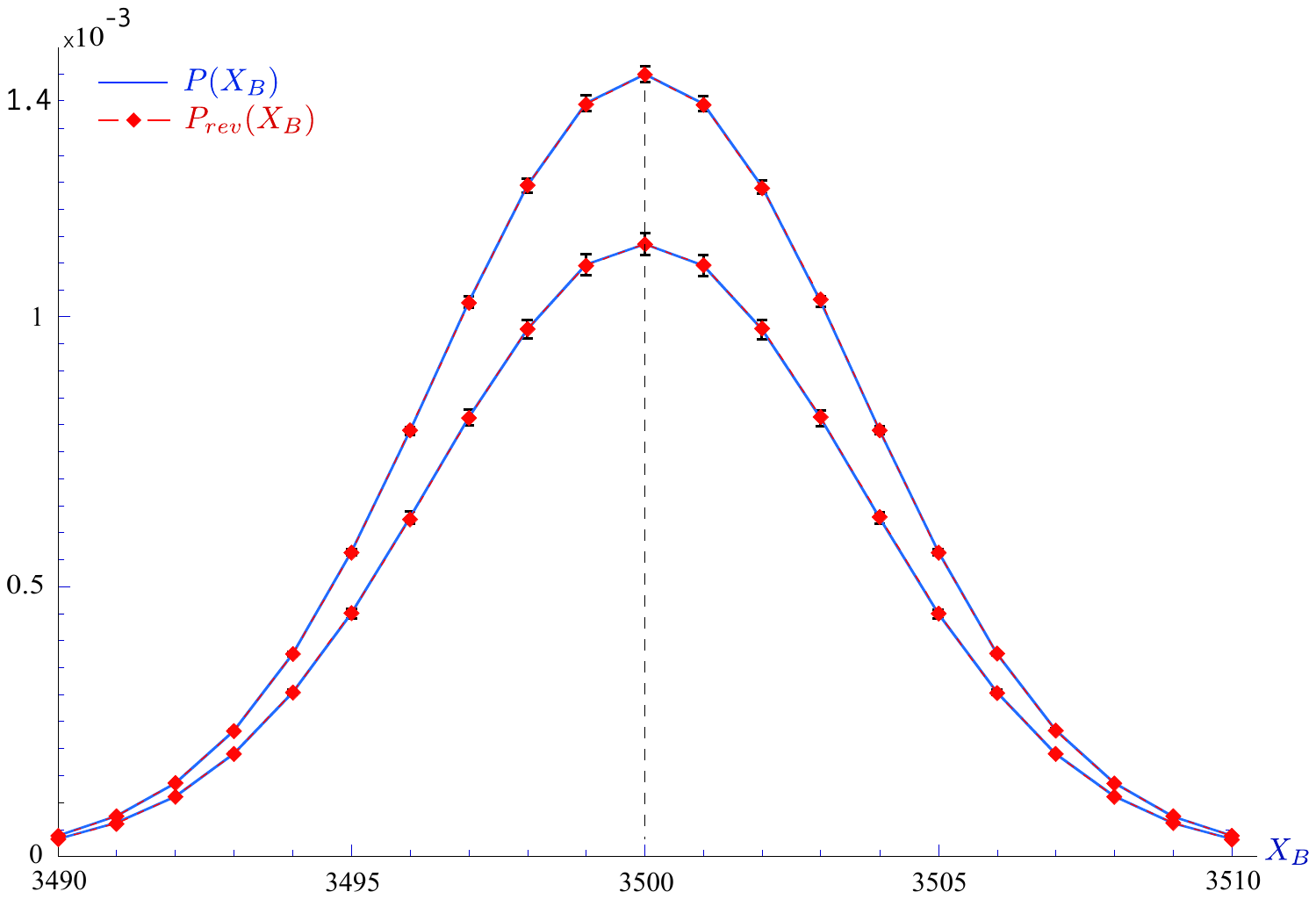,width=1.\textwidth} 
\caption{Direct probability distribution, $P(X_B, t; \, X_B^{\mathrm{ref}}, t + \tau)$ (blue line), and the corresponding reverse probability distribution, $P(X_B^{\mathrm{ref}}, t; \, X_B, t + \tau)$ (red line), for both scenarios I (upper curve) and II (lower curve).}
\label{Fig2}
\end{center}
\end{figure}

To perform the measurements, we consider a total number of $N = 7000$.  We then generated two independent data sets corresponding to scenarios I and II, each comprising simulations totaling $2.55 \times 10^{10}$ collisions. The corresponding direct and reverse probability distributions are shown in Figure~\ref{Fig2} for both scenarios I and II, with $\tau = 10$~MCT. The statistical error, estimated from $51$ successive simulations comprising $5 \times 10^{8}$ collisions each, does not exceed 1\% for scenario I and is slightly higher for scenario II.

Quite surprisingly, we observe that the direct and reverse trajectories are practically indistinguishable in both scenarios.  In other words, $P(X_B, t; \, X_B^{\mathrm{ref}}, t + \tau) \approx P(X_B^{\mathrm{ref}}, t; \, X_B, t + \tau)$, well within the limits of statistical errors, despite the fact that the system is \emph{not} at equilibrium.  

\begin{figure}[ht!]
\begin{center}
\epsfclipon
\epsfig{file=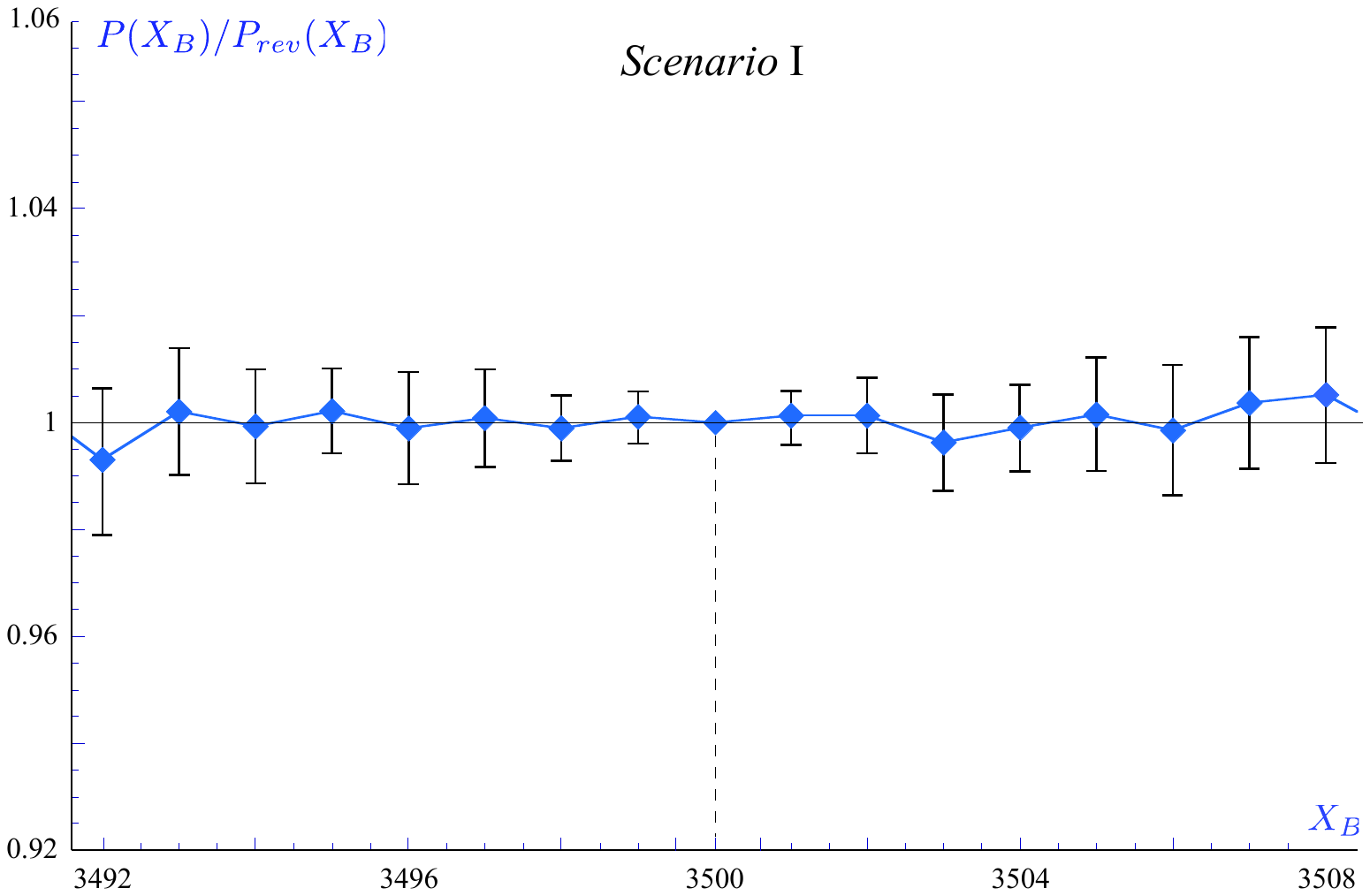,width=0.9\textwidth}
\caption{The ratio of the direct and reverse probability distributions, $P(X_B)/P_{rev}(X_B) = P(X_B, t; \, X_B^{\mathrm{ref}}, t + \tau)/P(X_B^{\mathrm{ref}}, t; \, X_B, t + \tau)$ for scenario I. }
\label{Fig3Big}
\end{center}
\end{figure}

This unexpected property can be further illustrated by examining the ratio of the direct and reverse probabilities. As shown in Figure~\ref{Fig3Big}, this ratio is essentially equal to unity in scenario I.  The same result is found in scenario II, albeit with a larger dispersion.

\section{Concluding remarks}\label{ConclusionsSection}
 
In Section~\ref{StochasticSection}, we presented a simple random-walk formulation of mass diffusion described by the so-called master equation \cite{N-P:1977}. From the exact solution of this equation, it can be shown that the sample paths of the total number of particles are time-reversible. To assess the validity of this unexpected result, we performed molecular dynamics simulations of the system in Section~\ref{MolDynSection}. Remarkably, the microscopic simulations lead to the same conclusion.
 
It is important to recall that, although the time-reversibility of state trajectories in non-equilibrium processes is a known phenomenon \cite{Malek:2020}, it has thus far been observed only for a particular class of reactive systems of the Schl\"{o}gl type, for which the underlying mechanism is well understood \cite{Baras:2023,Politano:2024}. To the best of our knowledge, no similarly simple explanation exists for the origin of the time-reversibility observed here in non-equilibrium fluid mixtures. Scientific integrity therefore compels us to acknowledge that, at present, we have not identified a satisfactory explanation for this phenomenon.

Another unexpected consequence of this result is that it implies that the entropy production evaluated along a stationary sample path of the system (path entropy production) is zero, even though the system is maintained out of equilibrium. This result appears to contradict the basic prediction of the well-known fluctuation theorem \cite{Evans:1993, Evans:1994}.
In this regard, it is useful to recall that the conventional entropy production, based on the Gibbs-Shannon definition of entropy, is proportional to the square of the concentration gradient and is therefore strictly positive in non-equilibrium systems \cite{Kondepudi:2008}.

Finally, one may speculate whether this unexpected result is specific to fluid mixtures subjected to a concentration gradient, or whether other simple systems might exhibit similar behavior when driven by an appropriate non-equilibrium constraint. In this context, a particularly relevant case that has recently attracted our attention is that of a solid subjected to a temperature gradient.

The principal advantage of considering a solid is that, under a thermal gradient, its macroscopic behavior is fully described by Fourier's law. However, the main difficulty is that molecular simulations of solids require substantially greater computational resources than those needed for dilute fluids. We are therefore currently investigating several types of microscopic simulation methods for dense materials in order to address this challenge.

\section*{Acknowledgments}
First of all, we would like to express our warmest thanks to Prof. J. W. Turner for his invaluable assistance in analyzing the behavior of the trajectories generated by the master equation (1).  The use of computational facilities at the Computing Center of the University of Bourgogne, DNUM-CCUB, is gratefully acknowledged.  One author (AG) acknowledges support by the U.S. Department of Energy, Office of Science, Office of Advanced Scientific Computing Research, Applied Mathematics Program under contract No. DE-AC02-05CH11231.

\end{document}